\documentclass[a4paper,fleqn,usenatbib]{mnras}

\usepackage{times}

\usepackage[T1]{fontenc}
\usepackage{ae,aecompl}

\usepackage{graphicx}	
\usepackage{amsmath}	
\usepackage{amssymb}	
\usepackage{epsfig}
\usepackage{pdflscape}

\def\msun{\hbox{M$_\odot$}}

\def\t4{\hbox{t$_{\rm 4}$}}

\def\cm3{\hbox{cm$^{-3}$}}

\input psfig.sty
\input epsf.sty

\title[Multiple populations in Hodge 6]{Spectroscopic detection of multiple populations in the $\sim$2~Gyr old cluster Hodge 6 in the LMC}

\author[Hollyhead et al.]{
K. Hollyhead$^{1}$\thanks{E-mail: kathie.hollyhead@astro.su.se},
S. Martocchia$^{2,3}$,
C. Lardo$^{4}$,
N. Bastian$^{3}$,
N. Kacharov$^{5}$,
\newauthor F. Niederhofer$^{6}$,
I. Cabrera-Ziri$^{7}$\thanks{Hubble fellow},
E. Dalessandro$^{8}$,
A. Mucciarelli$^{9}$,
M. Salaris$^{3}$ and
\newauthor C. Usher$^{3}$
\\
$^{1}$ Department of Astronomy, Oscar Klein Centre, Stockholm University, AlbaNova, Stockholm SE-106 91, Sweden\\
$^{2}$European Southern Observatory, Karl-Schwarzschild-Stra\ss{}e 2, D-85748 Garching bei M\"{u}nchen, Germany\\
$^{3}$Astrophysics Research Institute, Liverpool John Moores University, 146 Brownlow Hill, Liverpool L3 5RF, UK\\
$^{4}$Laboratoire d'astrophysique, Ecole Polytechnique F\`ed\`erale de Lausanne (EPFL), Observatoire de Sauverny, CH-1290 Versoix, Switzerland\\
$^{5}$Max-Planck-Institut f\"{u}r Astronomie, K\"{o}nigstuhl 17, D-69117 Heidelberg, Germany\\
$^{6}$Leibniz-Institut f\"{u}r Astrophysik Potsdam, An der Sternwarte 16, Potsdam 14482, Germany\\
$^{7}$Harvard-Smithsonian Center for Astrophysics, 60 Garden Street, Cambridge, MA 02138, USA\\
$^{8}$INAF, Osservatorio Astronomico di Bologna, via Ranzani 1, 40127, Bologna, Italy\\
$^{9}$Department of Physics and Astronomy, University of Bologna, Viale Berti Pichat 6/2, I-40127 Bologna, Italy\\
}

\date{Accepted XXX. Received YYY; in original form ZZZ}

\pubyear{2018}

\begin{document}
\label{firstpage}
\pagerange{\pageref{firstpage}--\pageref{lastpage}}
\maketitle

\begin{abstract}
We report the spectroscopic discovery of abundance spreads (i.e. multiple populations) in the $\approx$2~Gyr old cluster in the LMC, Hodge~6. We use low resolution VLT FORS2 spectra of 15 member stars in the cluster to measure their CN and CH band strengths at $\simeq$ 3883 and 4300~\AA, respectively, as well as [C/Fe] and [N/Fe] abundances. We find a sub-population of 2 stars that are enriched in nitrogen, and we conclude that this sub-population is evidence of multiple populations in Hodge~6. This is the second $\sim$2~Gyr old cluster (the first being NGC~1978 in the LMC) to show multiple populations and the first spectroscopic detection of MPs in a cluster of this age. This result is interesting as it hints at a possible relationship between the disappearance of extended main sequence turn-offs in clusters younger than $\approx$2~Gyr and the onset of multiple populations at $\approx$2~Gyr, which should be explored further.       
\end{abstract}




\section{Introduction}
\label{sec:intro}

Multiple populations \citep[MPs, chemical variations e.g. O-Na, C-N anti-correlations and splits/spreads in colour magnitude diagrams, which are ubiquitous to globular clusters, e.g.][]{mucciarelli09,carretta09,gratton12,piotto15} have recently been identified in intermediate age (2-8 Gyr) massive clusters in the SMC and LMC, spanning this full range of ages. These include Lindsay~1 \citep[$\approx$8~Gyr;][]{hollyhead17,niederhofer17}, Kron~3, NGC~416 and NGC~339 \citep[$\approx$6~Gyr old;][]{niederhofer17,hollyhead18}. Most recently, the $\approx$2~Gyr old cluster NGC~1978 was shown to have evidence of a split red giant branch (RGB) and sub-giant branch (SGB) in its colour magnitude diagram \citep[CMD,][]{martocchia18b,martocchia18a}.  

Clusters younger than $\approx$2~Gyr, though of comparable mass to the aforementioned SMC clusters and globular clusters (GCs), show a lack of evidence for MPs spectroscopically \citep[e.g. NGC~1806][]{mucciarelli14} and photometrically \citep[e.g. NGC~419][]{martocchia17}. Interestingly, open clusters of comparable ages and masses to clusters with MPs have been found to lack MPs in some cases \citep[e.g. Berkeley~39 and NGC~6791, masses $\sim$ 10 $^4$ \msun, ages 6~Gyr and 7-8~Gyr, respectively;][]{bragaglia12,bragaglia14,kassis97,platais11,brogaard12}, and show evidence in others \citep{pancino18}. Having sufficient mass is known to be a key factor in whether or not a cluster forms MPs \citep{carretta10,schiavon13,milone17}, with the lowest mass GCs that host MPs being 10$^{3.5 - 3.9}$\msun\ \citep{bragaglia17,milone17,simpson17}. However, the recent discoveries of MPs in intermediate age clusters ($\approx$ 2-8 Gyr), though not in young massive clusters (YMCs, < 2~Gyr old) indicate that age also plays a key role. The precise mechanism of the formation of MPs and its direct relationship to age is still not fully understood \citep[e.g.][]{bastianlardo17}. 

Despite a lack of evidence of MPs in YMCs spectroscopically, extended main sequence turn-offs (eMSTOs) have been observed in many of their CMDs \citep[e.g.][]{mackey07,piatti14}. This phenomenon is not observed in any clusters of any mass older than $\approx$2~Gyr. Though these observations were originally explained with an age spread in the cluster of 200-700~Myr \citep[e.g.][]{milone09}, it has since been shown that this is unlikely due to the width of the turn-off being proportional to the age of the cluster \citep{niederhofer15}, and the discovery of extended turn-offs in clusters younger than the possible age spreads \citep[e.g. NGC~1850, 100~Myr old,][]{bastian16}. Though it has been suggested that the spreads are related to the MP phenomenon as observed in GCs \citep[e.g][]{goudfrooij14}, the size of the spreads and lack of eMSTO after $\sim$2~Gyr is described well by invoking stellar rotation \citep[e.g.][]{bastian09,brandt15,dantona15}. 

These discoveries have consequences for GC formation theories, which try to explain how MPs have formed. All current theories \citep[e.g.][]{decressin07,dercole08,bastian13} cannot fully explain observations \citep{bastianlardo15,renzini15} and have significant issues \citep[e.g. mass-budget problem;][]{larsen12,kruijssen15}. The detection of MPs in clusters down to $\approx$2~Gyr, however, helps to constrain these theories as it suggests the process should still be operating in the present day universe. This also means that YMCs can be used to constrain these theories, as they likely formed through the same mechanism. Studies of YMCs suggest that MPs should be formed in a single burst of star formation, due to a lack of gas reservoir to form a second generation \citep[e.g.][]{ivan15}.  
 
The aim of this project is to further investigate clusters around the age of $\sim$2~Gyr, which marks the point where clusters no longer have eMSTOs and where MPs have been identified in NGC~1978 \citep{martocchia18a}. By increasing the sample of clusters at this age, the onset of MPs can be further constrained and the relationship between the loss of eMTSOs and the appearance of MPs can be explored. 

In this paper we analyse Hodge~6, a $\approx$2~Gyr cluster \citep{goudfrooij14} in the LMC. \cite{goudfrooij14} give the mass of Hodge~6 as 8$\times$10$^4$\msun\ using a Salpeter IMF. If a Kroupa/Chabrier IMF is used Hodge~6 is $\sim$5$\times$10$^4$\msun. This cluster was chosen for this study as it is shown to lack an eMSTO by \cite{goudfrooij14} and at $\approx$2~Gyr is at the limit where this phenomenon is observed, so can be used to explore the role this transition plays in the formation of MPs. 

We have obtained low resolution spectroscopy of lower RGB stars in the cluster in order to look for the signature of MPs in CN and CH band strengths \citep[enrichment in N, e.g.][]{norris81,cohen02,kayser08,pancino10,lardo12} that trace N and C, respectively. We also observed NGC~1978 on the same observing run, however after reduction and analysis, the data was deemed to be unusable, therefore we present only Hodge~6 in this paper. We discuss why NGC~1978 is not used and how we ensured that Hodge~6 data was still viable in \S~\ref{sec:contam}.

In \S~\ref{sec:data} we describe our data and briefly discuss the data reduction, while \S~\ref{sec:members} describes how we differentiate between cluster members and field interlopers. The calculation of the CN and CH band strengths is discussed in \S~\ref{sec:cnchoverall}, [C/Fe] and [N/Fe] calculations in \S~\ref{sec:cn}, along with a new age estimate in \S~\ref{sec:age}. Finally, the results and discussion are included in \S~\ref{sec:results}.

\section{Observations and data reduction}
\label{sec:data}

The data for Hodge~6 was obtained from FORS2 on the VLT at La Silla Paranal Observatory under the Programme ID 099.D-0762(B), P.I. K. Hollyhead. We used the same configuration as for previous observations for Lindsay~1 and Kron~3 \citep[multi-object spectroscopy using the blue CCD in visitor mode with the 600B+22 grism, in order to sample the CN and CH bands at 3839~\AA~ and 4300~\AA, respectively;][]{hollyhead17,hollyhead18}.  Three science exposures were obtained over the course of one half-night, covering 37 target stars across the two chips. 

Pre-imaging was taken prior to the spectra under the same programme, obtaining V and I band images, which were used to select appropriate targets for the spectroscopy, in addition to calculating properties of the stars. The images were reduced using the Reflex pipeline \citep{reflex}, and PSF photometry was used to obtain magnitudes for stars within the images with {\sc daophot} \citep{daophot}. 

Targets for spectroscopic follow-up were selected from the CMD resulting from the PSF photometry. The targets were chosen from the stars along the RGB, with priority targets as those on the lower RGB and secondary targets closer to the tip. When the masks were produced for Hodge~6, secondary targets were selected only in the absence of primary targets, and where neither of these were available, random stars were selected to fill gaps. 

\begin{figure}
	\includegraphics[width=8.5cm]{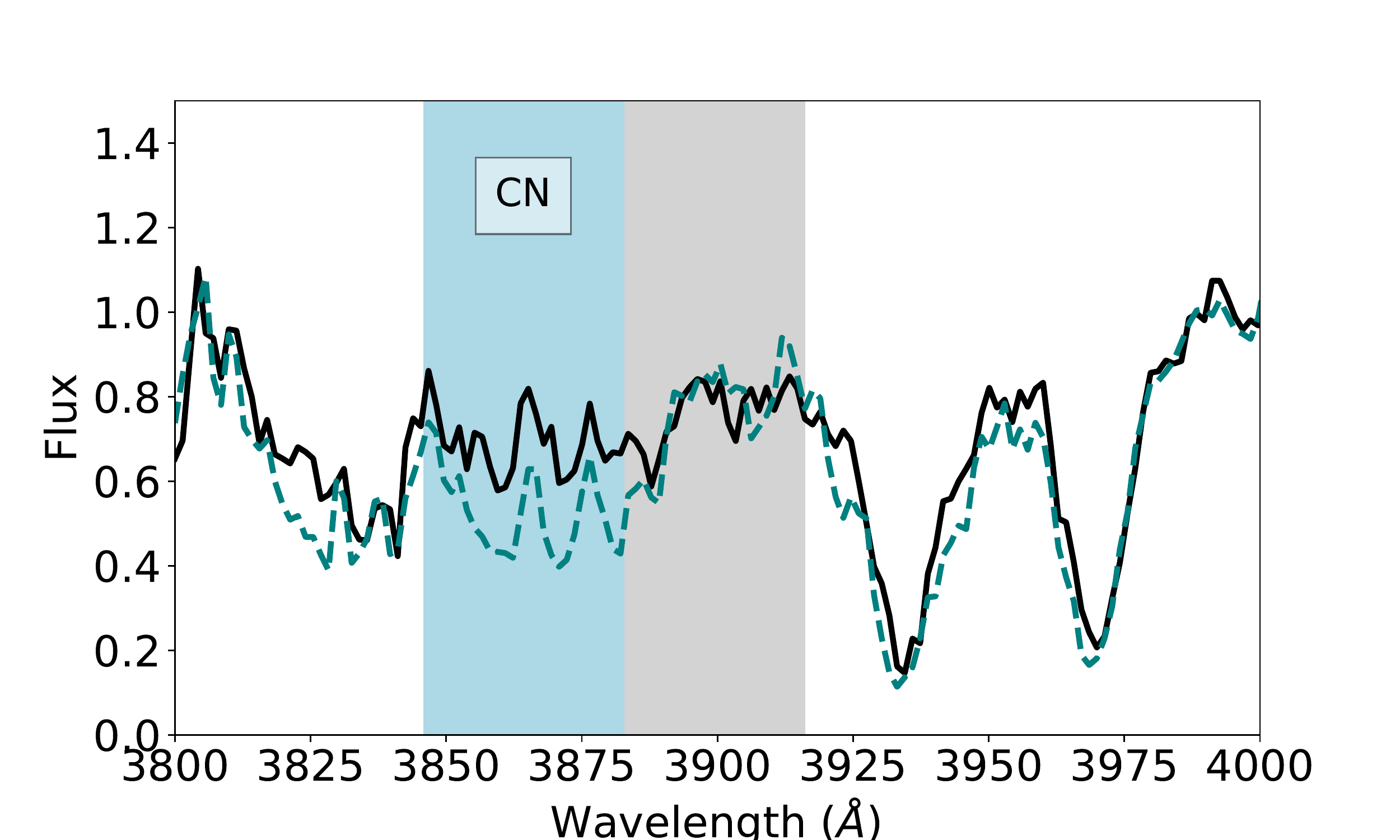}
	\includegraphics[width=8.5cm]{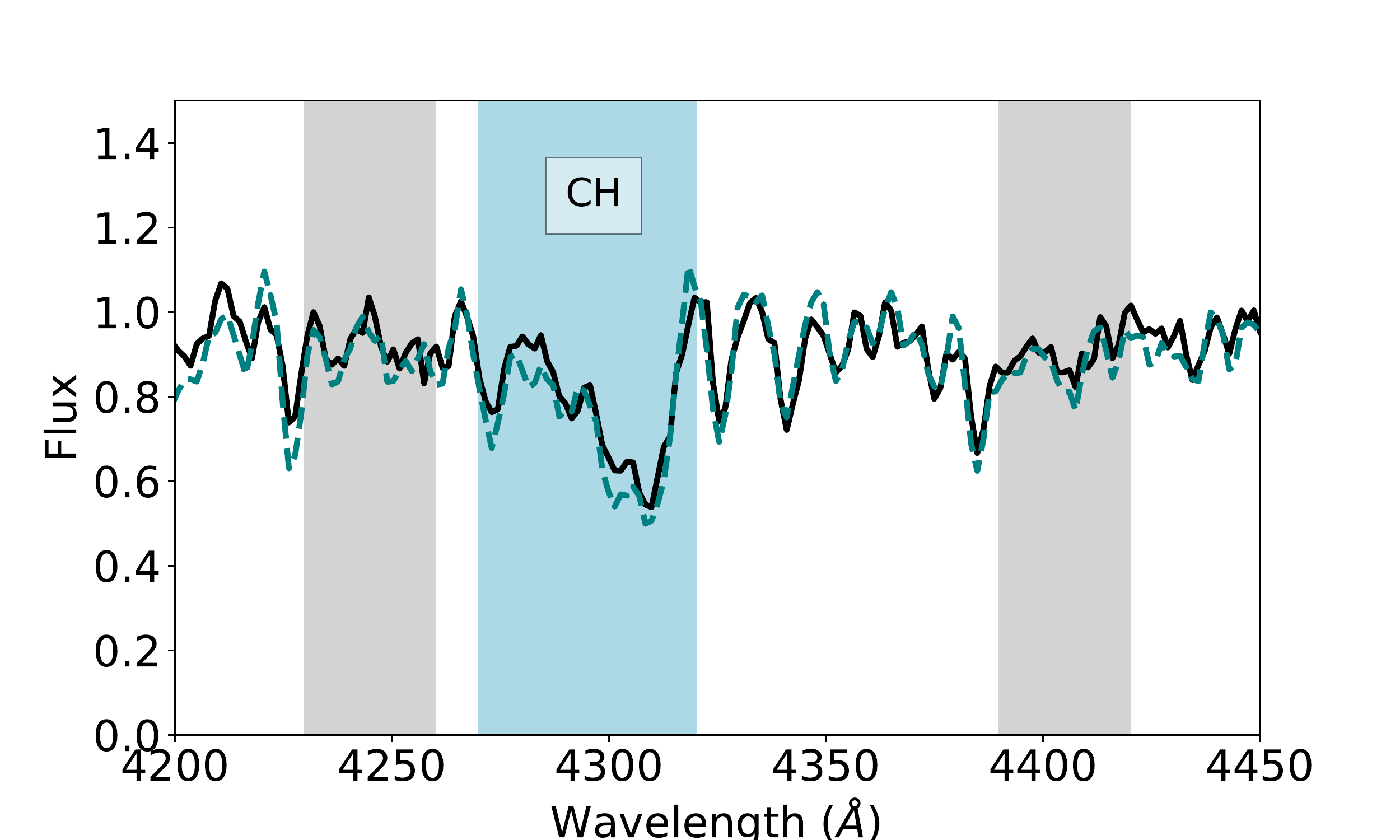}
	\caption{Examples of spectra in our sample. The spectra of a CN strong and a CN normal are shown in black and teal, respectively. They have both been continuum normalised using {\sc iraf}. In the top plot we show the wavelength range including the CN band (shown in blue) and its continuum band (shown in grey). The  lower plot shows the CH band. The two stars selected have similar properties (T$_{eff}$ and magnitude) though they show a clear difference in their CN band. }
	\label{fig:spectra}
\end{figure}

The spectra were reduced using {\sc iraf}. The spectra were bias subtracted, flat fielded with a normalised flat field image, cleaned of cosmic rays with the L.A.Cosmic \citep{lacos} routine and extracted. The 1D spectra were then wavelength calibrated and combined. The average S/N for our spectra is lower than for our previous studies, with averages of $\approx$ 11 and 16 for the CN and CH bands, respectively. Examples of the spectra are shown in Fig.~\ref{fig:spectra}.  

\section{Cluster membership}
\label{sec:members}

We applied a number of different criteria to determine cluster membership. The process was the same as applied to the targets in Lindsay~1 and Kron~3 previously in \cite{hollyhead17,hollyhead18}. 

\begin{figure}
	\includegraphics[width=8.6cm]{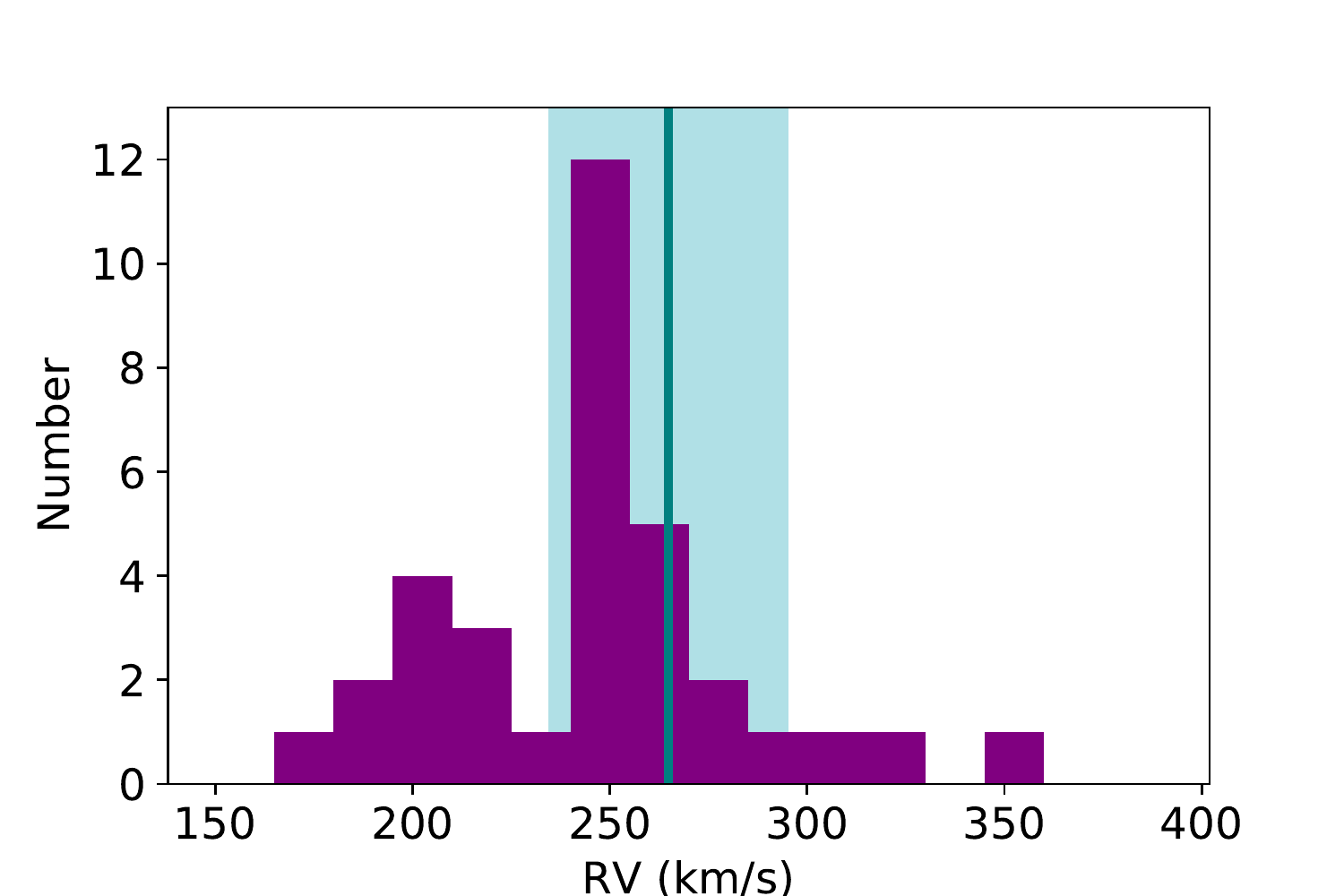}
	\caption{Histogram of the radial velocities of all stars with spectra in Hodge~6. The teal line shows the radial velocity found for our selected template star, indicating that it is very likely to be a member as it is close to the peak of RVs. The filled blue section shows the range over which we selected member stars.}
	\label{fig:rvhist}
\end{figure}

\begin{figure}
	\includegraphics[width=8.5cm]{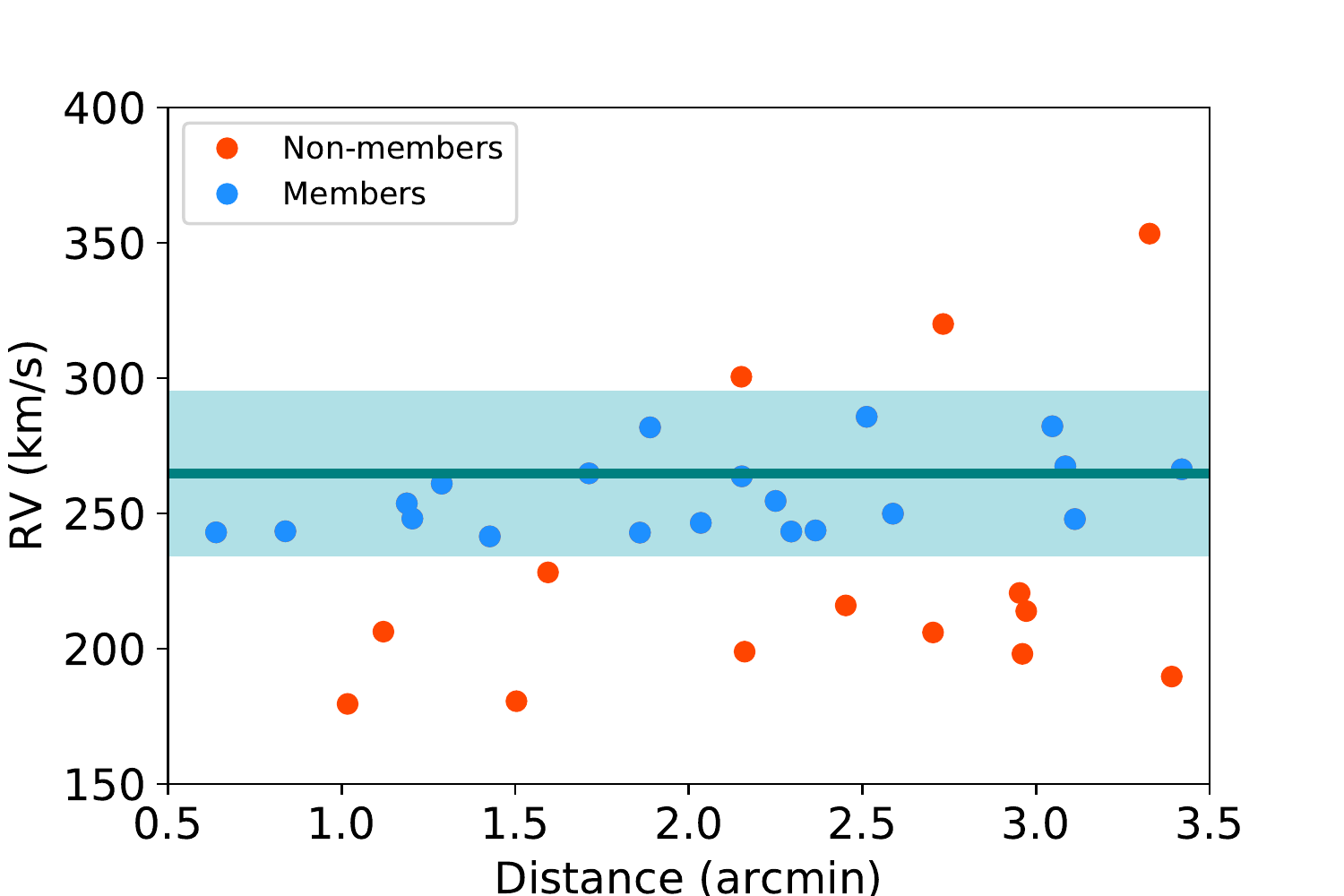}
	\includegraphics[width=8.5cm]{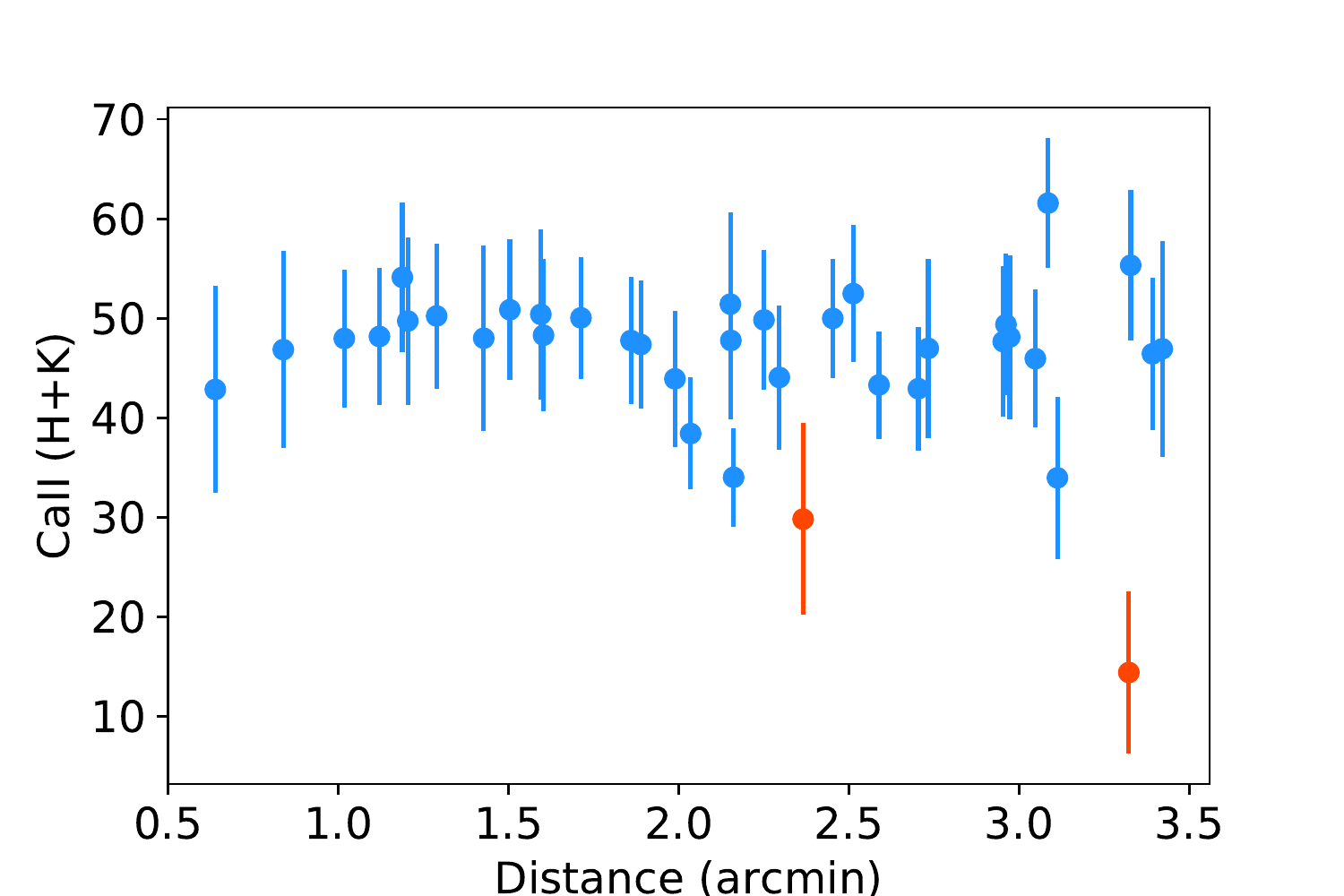}
	\caption{Here we show the criteria used to determine cluster membership for each of our stars. In all cases the blue stars were determined to be members from each test and red points are non-members. They were cross-correlated and any star failing any of the three criteria was removed. The top panel shows the radial velocities of the stars against their distance from the centre of the cluster, with the RV of the template star as the teal line and the 30~km/s range by the shaded blue. Stars outside of this range were considered non-members. The second panel shows the Ca{\sc ii} (H+K) estimates for each star from band strengths. Stars outside of 2$\sigma$ from the median were classified as non-members.} 
	\label{fig:members}
\end{figure}

Firstly a cut was made based on the radial velocities (RVs) of the targets. RVs were derived using {\sc iraf}. The star with the highest S/N was selected as the template spectrum for calculating the RVs of the other stars. The {\it rvidlines} routine was used to determine the RV of the template spectrum (star 11 on chip 1) as 264.8 km/s. The error on the RV measurements was estimated by measuring the RV of the template spectrum in {\it rvidlines} using lines at the bluer end of the spectrum compared to lines at the redder end as $\approx$ 30 km/s. Any stars more or less than 30 km/s from the template RV were removed as non-members. This is shown in the top panel in Fig.~\ref{fig:members} with each star's RV plotted against its distance from the centre of the cluster, and where the teal line is the template RV, and the blue shaded region indicates the acceptable range of velocities for member stars. Red points are likely non-members and blue points are members. 

In order to check that our template star was a member star itself, we looked at the histogram of the RVs, and star 11 was very close to the peak, and therefore likely a member. The histogram is shown in Fig.~\ref{fig:rvhist}. The teal line shows the RV of star 11 and the blue shaded area is the selection of RVs, as described above. Our selection well samples the peak velocities while also missing a secondary peak of likely non-member stars at $\sim$ 200 km/s. 

The second panel in Fig.~\ref{fig:members} shows estimates of the metallicity of each star, for which the strength of the Ca{\sc ii} (H+K) lines is a proxy. Once again, the blue points are members and red points are non-members. Member stars were selected as those less than 2$\sigma$ from the median of all stars, the same cut as applied for Kron~3 \citep{hollyhead18}. Previously we have used the Fe5270 band strengths, which are also proxies for metallicity, as a further criterion for membership. However, in this case no outliers could be determined due to large errors, so it is not used for Hodge~6. 


\begin{figure}
	\includegraphics[width=8.5cm]{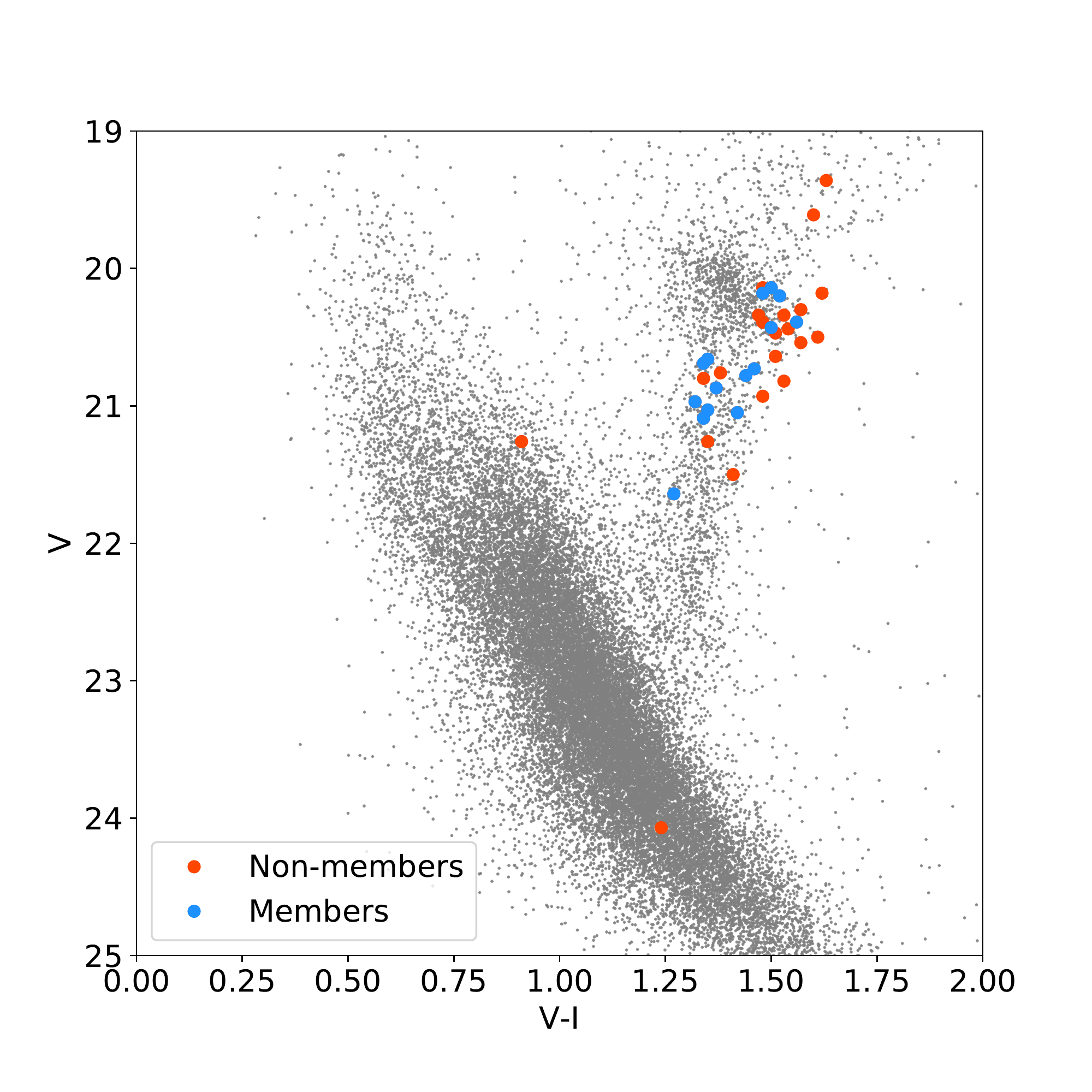}
	\caption{CMD of all the stars in the pre-imaging field for Hodge~6, with photometry obtained using {\sc daophot} \protect\citep{daophot}. The grey points are all stars in the photometry, while member stars with spectroscopy are shown as blue points and non-members (determined from the CMD and other sources) are shown as red points.  }
	\label{fig:cmd}
\end{figure}   

The CMD of the targets was then inspected to ensure no random filler stars were still considered members of the cluster. Fig.~\ref{fig:cmd} shows the CMD using the photometry obtained from pre-imaging of the cluster. Grey points are all the sources within the catalogue, red points are the non-members and blue points are members. Stars brighter than the bump were excluded from the final sample as the C and N abundances are significantly changed by internal mixing processes. Additionally several stars were too far to the right of the RGB and were removed. After applying all of these criteria, we were left with 15 member stars out of the total 37. All RVs and HK band strengths are listed in Table~\ref{tab:results}. 

\subsection{Bright star contamination}
\label{sec:contam}

As mentioned previously, spectroscopy was also obtained for NGC~1978. This data could have been used as a further check of our method for determining MPs, as they have been shown to be present in HST photometry. However, during the reduction of the data we discovered that most of the member stars' spectra were contaminated by nearby bright stars, therefore making the data unreliable and unusable for this study.

After this was found, we checked the spectra for Hodge~6 to ensure that the data for this cluster was viable. By examining the the spectra, images and the brightness and colour of all sources within 5 arcseconds, we determined that 10 of our 15 member stars were uncontaminated or had only very faint stars nearby. We also checked HST ACS images of the cluster for a more thorough check of any stars within the field of view. 

\section{Band strengths and abundances}
\label{sec:cnchoverall}

\subsection{CN and CH band strengths}
\label{sec:cnch}

We have calculated the band strengths S$\lambda$3883 (CN) and CH$\lambda$4300 (CH) to investigate the presence of multiple populations in Hodge~6. This technique has been used previously for Milky Way Globular Clusters \citep[e.g.][]{pancino10} and for our other intermediate age clusters Lindsay~1 and Kron~3 \citep{hollyhead17,hollyhead18}. 

The band strengths are calculated using the definitions in \cite{norris81}, \cite{Worthey94} and \cite{Lardo13}. Errors on each measurement are calculated as per \cite{Vollmann06}. The bands used for the calculations, including the ranges used as estimates of the continuum for each band are shown in Fig.~\ref{fig:spectra}, with blue as the molecular band and grey as the continuum bands. Our values for the CN and CH band strengths for each star are listed in Table~\ref{tab:results}.  

\subsection{[C/Fe] and [N/Fe]}
\label{sec:cn}

We also derived [C/Fe] and [N/Fe] abundances for the 10 uncontaminated member stars using the same methods as for previous clusters Lindsay~1 and Kron~3. We carried out spectral synthesis using the same bands as for the band strength determinations to obtain estimates of [C/Fe] and [N/Fe]. 
	
We used Kurucz line lists taken from the website of F. Castelli\footnote{\url{http://wwwuser.oats.inaf.it/castelli/linelists.html}} and used ATLAS9 to produce model atmospheres using [Fe/H]=-0.3 \citep{goudfrooij14} and the parameters derived for each star. Effective temperture was found using a $T_{\rm{eff}}$-colour calibration \citep{Alonso99} with V-I, using our pre-imaging photometry. Surface gravities were calculated using a distance modulus of 18.4 \citep{goudfrooij14}, the previously determined effective temperatures and bolometric corrections also from \cite{Alonso99}. 

Kurucz's {\sc synthe} code was used to create model spectra, which were used in a $\chi^{2}$ minimisation with the observed spectra to find the abundances. Measured [C/Fe] abundances were used in finding [N/Fe] across the CN band, with solar abundances taken from \cite{Asplund09}. In order to calculate uncertainties from the fitted parameters, we iteratively change one parameter and repeat the abundance analyses for the full range of temperatures and gravities, as used in \cite{Lardo13}. Error introduced from the  $\chi^{2}$ fitting procedure (found by re-fitting after introducing Poissonian noise) is added in quadrature to give the final errors in Table~\ref{tab:results}. We find very large errors due to the noise in the spectra at the bands used, and the difficulty in calculating [C/Fe] and [N/Fe] using molecular bands, rather than individual lines.

\section{The age of Hodge 6}
\label{sec:age}

\begin{figure}
	\includegraphics[width=8.5cm]{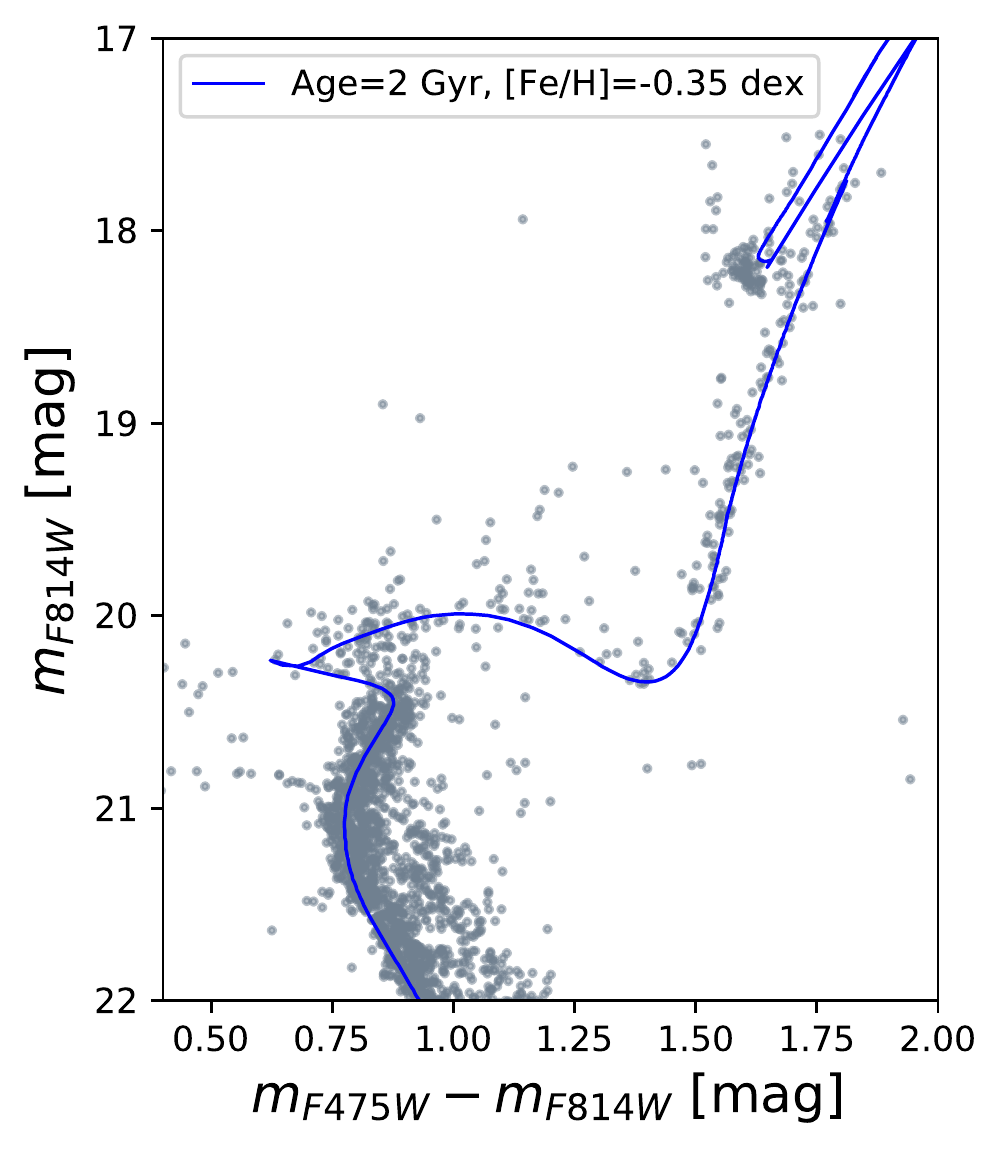}
	\caption{CMD for Hodge~6 constructed with ACS photometry in F475W and F814W filters. The CMD has been field star subtracted. The best fit BaSTI isochrone is shown, giving Hodge~6 an age of $\approx$2~Gyr and [Fe/H] of -0.35~dex. }
	\label{fig:agefit}
\end{figure}

In order to more accurately determine the age of Hodge~6, which is not greatly discussed in the literature, we fit a CMD produced with existing photometry of Hodge~6 \citep{goudfrooij14}, which was field star subtracted as per \cite{niederhofer17}. Fig.~\ref{fig:agefit} shows the CMD in the F475W and F814W filters with the best fit BaSTI isochrone \citep[parameters age = 2~Gyr, \text{[Fe/H]} = -0.35 dex, distance modulus = 18.5 and Av = 0.22,][]{basti}, giving Hodge~6 an age of $\approx$2~Gyr.   

We have also directly compared the CMD of Hodge~6 with that of NGC~1978, though the blue filter differs between the two observations (i.e., F475W vs. F555W).  Comparing the two (with F814W on the y-axis, which is common between the two) led us to conclude that the two clusters were coeval within $\sim$200 Myr.  As discussed previously, \cite{martocchia18b,martocchia18a} found MPs within NGC~1978 and estimated an age of $\sim$2 Gyr based on isochrone fitting. 

\section{Results and discussion}
\label{sec:results}

\begin{landscape}
\begin{table}
	\begin{tabular}{cccccccccccccc}
	ID & RA & Dec & V (mag) & I (mag) & RV (km/s) & HK (mag) & CN & CH & [C/Fe] & [N/Fe] & Member & Cont & Enriched\\
	\hline
	15219	&	 85.5362524 & 	-71.5351338  &   21.09  $\pm$ 0.02 & 19.75 $\pm$ 0.01 &  266 &  46.9 $\pm$ 10.8 & -0.03 $\pm$ 0.07 & -0.32 $\pm$ 0.12 & -0.34 $\pm$ 0.22 & -0.48 ul         & 1 & 0   & 0 \\
	9794	&	 85.6095721 &	-71.5429143  &   20.69  $\pm$ 0.02 & 19.35 $\pm$ 0.01 &  214 &  48.1 $\pm$ 8.2  & 0.15  $\pm$ 0.08 & -0.39 $\pm$ 0.08 &	      --- 		 &     --- 	        & 0 & N/A & N/A \\
	8311	&	 85.6300597 &	-71.5533655  &   20.14  $\pm$ 0.01 & 18.66 $\pm$ 0.01 &  286 &  52.5 $\pm$ 6.9  & 0.42  $\pm$ 0.08 & -0.35 $\pm$ 0.1  &	-0.45 $\pm$ 0.20 & 0.34 $\pm$ 0.27  & 1 & 0   & 1 \\
	11665	&	 85.5840111 &	-71.5553459  &   20.34  $\pm$ 0.01 & 18.87 $\pm$ 0.01 &  301 &  51.4 $\pm$ 9.3  & 0.13  $\pm$ 0.06 & -0.36 $\pm$ 0.08 &	      --- 		 &     --- 	        & 0 & N/A & N/A \\   
	9522	&	 85.6134158 &	-71.5575938  &   20.43  $\pm$ 0.01 & 18.93 $\pm$ 0.01 &  264 &  47.8 $\pm$ 7.9  & -0.05 $\pm$ 0.05 & -0.39 $\pm$ 0.08 &	-0.33 $\pm$ 0.20 & -0.70 ul         & 1 & 0   & 0 \\
	14926	&	 85.5403979 &	-71.5642641  &   20.20  $\pm$ 0.01 & 18.68 $\pm$ 0.01 &  265 &  50.0 $\pm$ 6.1  & 0.13  $\pm$ 0.08 & -0.36 $\pm$ 0.08 &	-0.39 $\pm$ 0.20 & 0.00 $\pm$ 0.27	& 1 & 0   & 0 \\
	8490	&	 85.6274101 &	-71.5710060  &	 20.14  $\pm$ 0.01 & 18.64 $\pm$ 0.01 &  228 &  50.4 $\pm$ 8.5  & 0.17  $\pm$ 0.09 & -0.32 $\pm$ 0.12 &	      --- 		 &     --- 	        & 0 & N/A & N/A \\
	17152	&	 85.5095419 &	-71.5664829  &   20.39  $\pm$ 0.01 & 18.91 $\pm$ 0.01 &  282 &  47.4 $\pm$ 6.4  & 0.08  $\pm$ 0.06 & -0.38 $\pm$ 0.09 &	-0.44 $\pm$ 0.20 & -0.32 $\pm$ 0.27 & 1 & 0   & 0 \\
	20569	&	 85.4611624 &	-71.5775750  &	 20.18  $\pm$ 0.01 & 18.70 $\pm$ 0.01 &  255 &  49.8 $\pm$ 7.0  & 0.25  $\pm$ 0.07 & -0.36 $\pm$ 0.08 &	-0.24 $\pm$ 0.19 & 0.12 $\pm$ 0.27  & 1 & 0   & 1 \\
	9065	&	 85.6197799 &	-71.5757263  &   21.05  $\pm$ 0.01 & 19.63 $\pm$ 0.01 &  261 &  50.2 $\pm$ 7.3  & 0.33  $\pm$ 0.13 & -0.38 $\pm$ 0.11 &	      --- 		 &     --- 	        & 1 & 1   & 1 \\
	8131	&	 85.6328257 &	-71.5852180  &	 20.78  $\pm$ 0.01 & 19.34 $\pm$ 0.01 &  248 &  49.7 $\pm$ 8.4  & 0.18  $\pm$ 0.08 & -0.39 $\pm$ 0.11 &	      --- 		 &     --- 	        & 1 & 1   & 0 \\
	8027	&	 85.6343877 &	-71.5890848  &   20.30  $\pm$ 0.01 & 18.73 $\pm$ 0.01 &  254 &  54.1 $\pm$ 7.5  & 0.47  $\pm$ 0.11 & -0.37 $\pm$ 0.1  &	      --- 		 &     --- 	        & 0 & N/A & N/A \\
	7152	&	 85.6470162 &	-71.5930918  &   20.73  $\pm$ 0.01 & 19.27 $\pm$ 0.01 &  242 &  48.0 $\pm$ 9.3  & 0.08  $\pm$ 0.11 & -0.39 $\pm$ 0.09 &	-0.48 $\pm$ 0.20 & -0.07 $\pm$ 0.27 & 1 & 0   & 0 \\
	9461	&	 85.6142583 &	-71.5951287  &   21.03  $\pm$ 0.01 & 19.68 $\pm$ 0.01 &  243 &  46.8 $\pm$ 9.9  & 0.1   $\pm$ 0.1  & -0.38 $\pm$ 0.11 &	      --- 		 &     --- 	        & 1 & 1   & 0 \\
	13362	&	 85.5616156 &	-71.5808774  &   20.97  $\pm$ 0.01 & 19.65 $\pm$ 0.01 &  243 &  42.8 $\pm$ 10.4 & 0.01  $\pm$ 0.08 & -0.46 $\pm$ 0.1  &	-0.58 $\pm$ 0.22 & -0.18 $\pm$ 0.30 & 1 & 0   & 0 \\
	12838	&	 85.5688218 &	-71.5478795  &   20.66  $\pm$ 0.01 & 19.31 $\pm$ 0.01 &  250 &  43.3 $\pm$ 5.4  & 0.07  $\pm$ 0.08 & -0.39 $\pm$ 0.09 &	-0.58 $\pm$ 0.22 & -0.25 $\pm$ 0.30 & 1 & 0   & 0 \\
	19789	&	 85.4722555 &	-71.5511889  &   20.39  $\pm$ 0.01 & 18.91 $\pm$ 0.01 &  282 &  45.9 $\pm$ 6.9  & 0.02  $\pm$ 0.06 & -0.38 $\pm$ 0.07 &	-0.27 $\pm$ 0.19 & -0.09 $\pm$ 0.27 & 1 & 0   & 0 \\
	12852	&	 85.5686579 &	-71.5454685  &   21.26  $\pm$ 0.01 & 19.91 $\pm$ 0.01 &  320 &  47.0 $\pm$ 9.0  & 0.1   $\pm$ 0.08 & -0.39 $\pm$ 0.1  &	      --- 		 &     --- 	        & 0 & N/A & N/A \\
	6987	&	 85.6495882 &	-71.5675732  &   21.26  $\pm$ 0.01 & 20.35 $\pm$ 0.01 &  247 &  38.4 $\pm$ 5.6  & -0.1  $\pm$ 0.03 & -0.49 $\pm$ 0.06 &	      --- 		 &     --- 	        & 0 & N/A & N/A \\
	10063	&	 85.6056276 &	-71.5407052  &   19.36  $\pm$ 0.01 & 17.73 $\pm$ 0.01 &  267 &  61.6 $\pm$ 6.5  & 0.55  $\pm$ 0.09 & -0.34 $\pm$ 0.12 &	      --- 		 &     --- 	        & 0 & N/A & N/A \\
	6192	&	 85.6125426 &	-71.6045965  &   20.34  $\pm$ 0.02 & 18.81 $\pm$ 0.01 &  206 &  48.2 $\pm$ 6.9  & 0.19  $\pm$ 0.07 & -0.39 $\pm$ 0.09 &	      --- 		 &     --- 	        & 0 & N/A & N/A \\
	7847	&	 85.5763716 &	-71.6078986  &   20.82  $\pm$ 0.02 & 19.29 $\pm$ 0.01 &  180 &  48.0 $\pm$ 6.9  & 0.26  $\pm$ 0.1  & -0.36 $\pm$ 0.11 &	      --- 		 &     --- 	        & 0 & N/A & N/A \\
	4211	&	 85.6575364 &	-71.6102169  &   20.64  $\pm$ 0.02 & 19.13 $\pm$ 0.01 &  -12 &  43.9 $\pm$ 6.9  & -0.01 $\pm$ 0.04 & -0.42 $\pm$ 0.07 &	      --- 		 &     --- 	        & 0 & N/A & N/A \\
	8384	&	 85.5650453 &	-71.6159635  &   20.50  $\pm$ 0.02 & 18.89 $\pm$ 0.01 &  181 &  50.9 $\pm$ 7.1  & 0.32  $\pm$ 0.1  & -0.36 $\pm$ 0.1  &	      --- 		 &     --- 	        & 0 & N/A & N/A \\
	9409	&	 85.5426523 &	-71.6205717  &   20.54  $\pm$ 0.02 & 18.97 $\pm$ 0.01 &  243 &  47.8 $\pm$ 6.4  & 0.14  $\pm$ 0.09 & -0.35 $\pm$ 0.1  &	      --- 		 &     --- 	        & 0 & N/A & N/A \\
	3662	&	 85.6704510 &	-71.6236386  &   20.44  $\pm$ 0.02 & 18.90 $\pm$ 0.01 &  206 &  42.9 $\pm$ 6.2  & 0.16  $\pm$ 0.06 & -0.37 $\pm$ 0.1  &	      --- 		 &     --- 	        & 0 & N/A & N/A \\
	10665	&	 85.5147734 &	-71.6260383  &   21.50  $\pm$ 0.02 & 20.09 $\pm$ 0.01 &  244 &  29.8 $\pm$ 9.6  & -0.17 $\pm$ 0.05 & -0.49 $\pm$ 0.08 &	      --- 		 &     --- 	        & 0 & N/A & N/A \\
	4558	&	 85.6497866 &	-71.6338156  &   20.47  $\pm$ 0.02 & 18.96 $\pm$ 0.01 &  198 &  49.4 $\pm$ 7.1  & 0.23  $\pm$ 0.05 & -0.37 $\pm$ 0.09 &	      --- 		 &     --- 	        & 0 & N/A & N/A \\
	7553	&	 85.5829425 &	-71.6290964  &   20.87  $\pm$ 0.02 & 19.50 $\pm$ 0.01 &  243 &  44.1 $\pm$ 7.2  & 0.07  $\pm$ 0.06 & -0.44 $\pm$ 0.09 &	      --- 		 &     --- 	        & 1 & 1   & 0 \\
	9237	&	 85.5462846 &	-71.6395408  &   20.39  $\pm$ 0.02 & 18.83 $\pm$ 0.01 &  221 &  47.7 $\pm$ 7.6  & 0.15  $\pm$ 0.07 & -0.38 $\pm$ 0.1  &	      --- 		 &     --- 	        & 0 & N/A & N/A \\
	10758	&	 85.5129849 &	-71.6443645  &   20.76  $\pm$ 0.02 & 19.38 $\pm$ 0.01 &  190 &  46.4 $\pm$ 7.7  & 0.36  $\pm$ 0.08 & -0.36 $\pm$ 0.1  &	      --- 		 &     --- 	        & 0 & N/A & N/A \\
	7246	&	 85.5892943 &	-71.6315025  &   20.18  $\pm$ 0.03 & 18.56 $\pm$ 0.01 &  216 &  50.0 $\pm$ 6.0  & 0.29  $\pm$ 0.1  & -0.37 $\pm$ 0.11 &	      --- 		 &     --- 	        & 0 & N/A & N/A \\
	10150	&	 85.5264974 &	-71.6135182  &   20.93  $\pm$ 0.02 & 19.45 $\pm$ 0.01 &  136 &  48.3 $\pm$ 7.6  & 0.11  $\pm$ 0.08 & -0.38 $\pm$ 0.1  &	      --- 		 &     --- 	        & 0 & N/A & N/A \\
	4671	&	 85.6471151 &	-71.6181124  &   19.61  $\pm$ 0.02 & 18.01 $\pm$ 0.01 &  199 &  34.0 $\pm$ 5.0  & -0.11 $\pm$ 0.03 & -0.43 $\pm$ 0.05 &	      --- 		 &     --- 	        & 0 & N/A & N/A \\
	3581	&	 85.6724476 &	-71.6365114  &   20.80  $\pm$ 0.02 & 19.46 $\pm$ 0.01 &  353 &  55.3 $\pm$ 7.6  & 0.36  $\pm$ 0.08 & -0.35 $\pm$ 0.11 &	      --- 		 &     --- 	        & 0 & N/A & N/A \\
	9607	&	 85.5384326 &	-71.6417748  &   21.64  $\pm$ 0.03 & 20.37 $\pm$ 0.01 &  248 &  34.0 $\pm$ 8.2  & -0.16 $\pm$ 0.06 & -0.48 $\pm$ 0.07 &	      --- 		 &     --- 	        & 1 & 1   & 0 \\
	8419	&	 85.5580955 &	-71.6461893  &   24.07  $\pm$ 0.05 & 22.83 $\pm$ 0.02 & 2709 &  14.4 $\pm$ 8.1  & -0.32 $\pm$ 0.03 & -0.55 $\pm$ 0.06 &	      --- 		 &     --- 	        & 0 & N/A & N/A \\
		\end{tabular}
	\caption{Properties of all the stars in our data set. ID refers to the ID for the star within our complete photometric catalogue. RVs are those calculated as in \S~\ref{sec:members}, HK refers to the Ca{\sc ii} (H+K) metallicity proxy, CN and CH are our band strengths, [C/Fe] and [N/Fe] are abundances, and 'Member' refers to which stars we found to be cluster members and non-members. A 1 indicates that the star is a member, 0 indicates that it is a non-member. 'Cont' refers to whether the star was found to be contaminated by bright sources or not, a 1 is contaminated and a 0 is not contaminated. Finally, 'Enriched' indicates whether the star has N enrichment or not. 1 means the star is enriched, 0 means it is non-enriched.}
	\label{tab:results}
\end{table}
\end{landscape}

\begin{figure}
	\includegraphics[width=8.5cm]{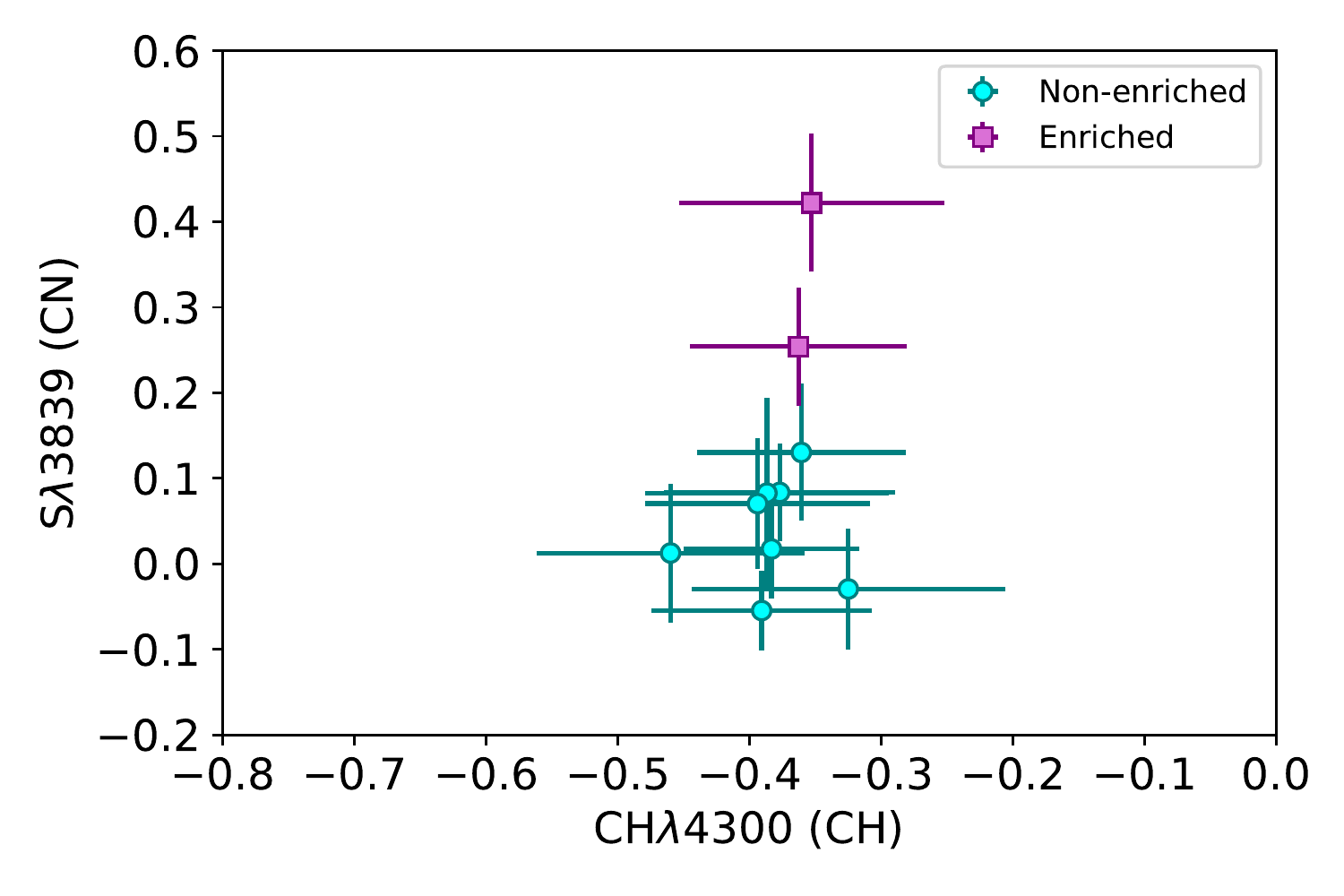}
	\caption{CN vs CH for all member stars in Hodge~6. A spread of $\sim$0.7 mag in N is much larger than the negligible spread in C. Purple points indicate the enriched stars, more than 1$\sigma$ from the median of the population. The blue points are non-enriched stars.}
	\label{fig:cnch} 
\end{figure}

In Fig.~\ref{fig:cnch} we show the distribution of CN and CH band strengths for all non-contaminated member stars, as listed in Table~\ref{tab:results}. The stars show a similar result to that observed for Kron~3 and Lindsay~1; a spread in CN, which traces nitrogen, ($\sim$0.7 mag) with negligible spread in CH, which traces carbon. We interpret this result to be evidence for multiple populations in Hodge~6. The blue points are the non-enriched stars and the three purple points indicate the N-enriched subpopulation, which we identify as more than 1$\sigma$ from the median of the CN distribution.    

\begin{figure*}
	\includegraphics[width=8.5cm]{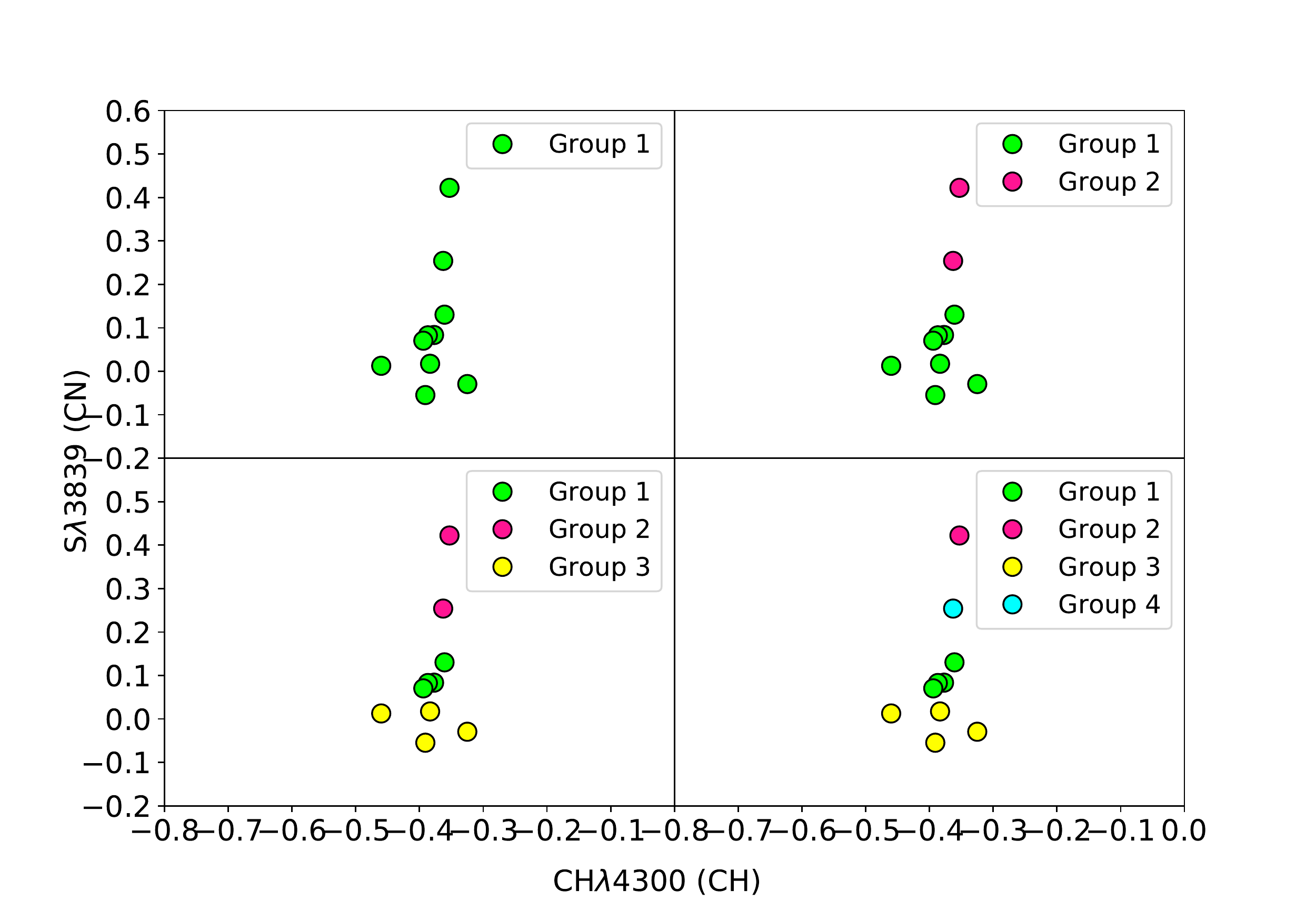}
	\includegraphics[width=8.5cm]{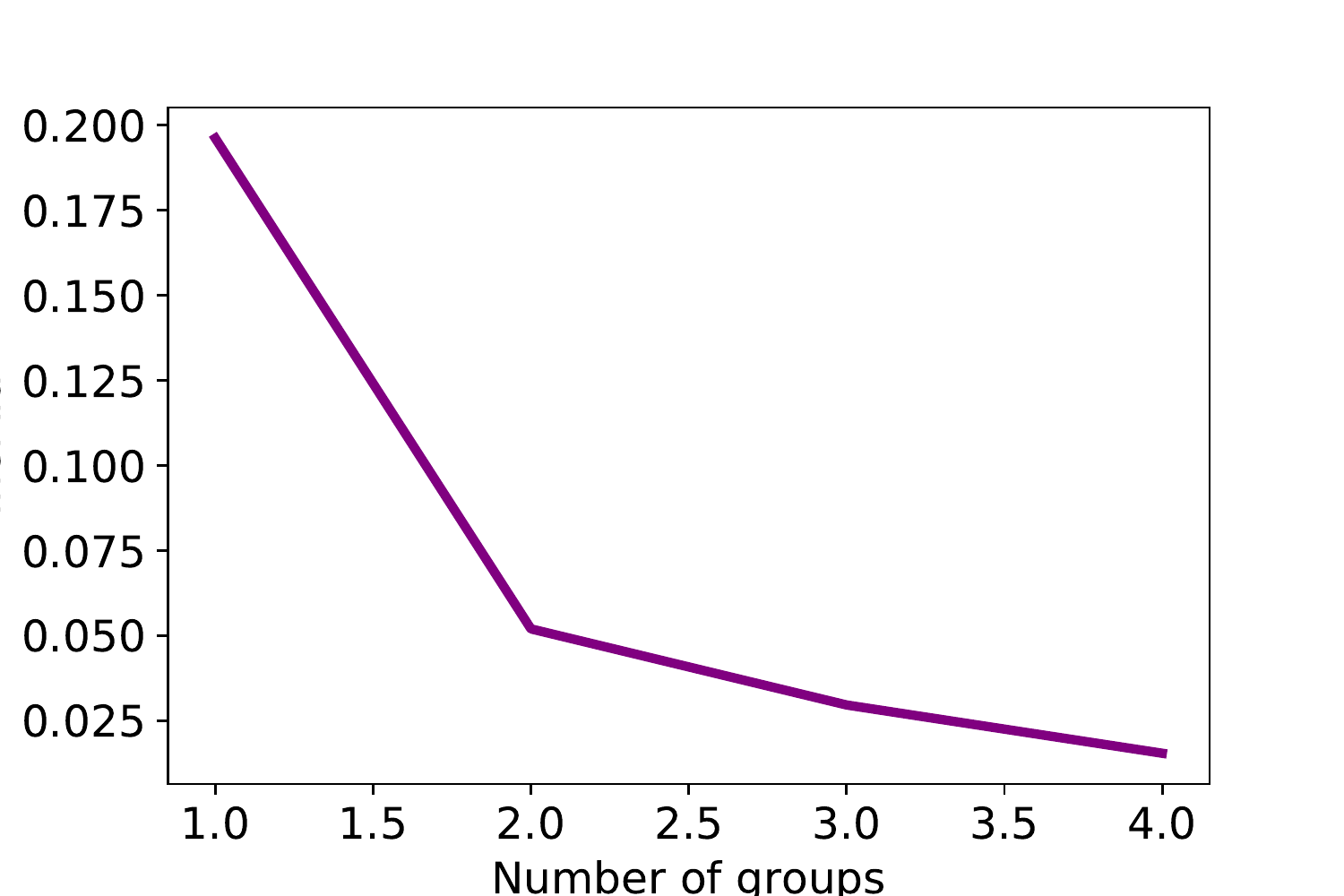}
	\caption{Results of the KMeans clustering algorithm ran on our data. The left plot shows the identified groups for inputs of 1-4 clusters, while the right plot shows the average distance between each group member and their cluster centre (inertia), used to find the optimal number of clusters.}
	\label{fig:clustering}
\end{figure*}

To further test the distribution of stars in CN/CH space and confirm that the three enriched stars are a separate population, we also ran a KMeans clustering algorithm on the data set using the {\sc sklearn} package in {\sc python}. In the left plot in Fig.~\ref{fig:clustering} we show the results of running the algorithm with 1-4 numbers of clusters. When a two cluster solution is requested, the algorithm selects the same three CN enhanced stars as a separate group to the rest, in agreement with our initial selection. However, as the number of clusters is selected, this does not mean that 2 is the optimal number of clusters for the data. To check this, we use the elbow test, where we plot the inertia (the average distance of each point to their cluster centre) against the number of clusters, as shown in the right plot in Fig.~\ref{fig:clustering}. The largest change in gradient between points indicates the best option for the number of clusters, in this case 2. 

The spectra in Fig.~\ref{fig:spectra} also clearly show the differences between spectra of CN-enhanced and non-enriched stars. We selected one CN-enriched star and one non-enriched star with very similar properties (i.e. magnitude) for the plot to show the difference in their CN band. The figure clearly shows that the enriched star in teal has enhanced CN, whereas both stars have very similar absorption in the CH band, in agreement with Fig.~\ref{fig:cnch} that shows a spread in CN with no corresponding spread in CH. 

\begin{figure*}
	\includegraphics[width=8.5cm]{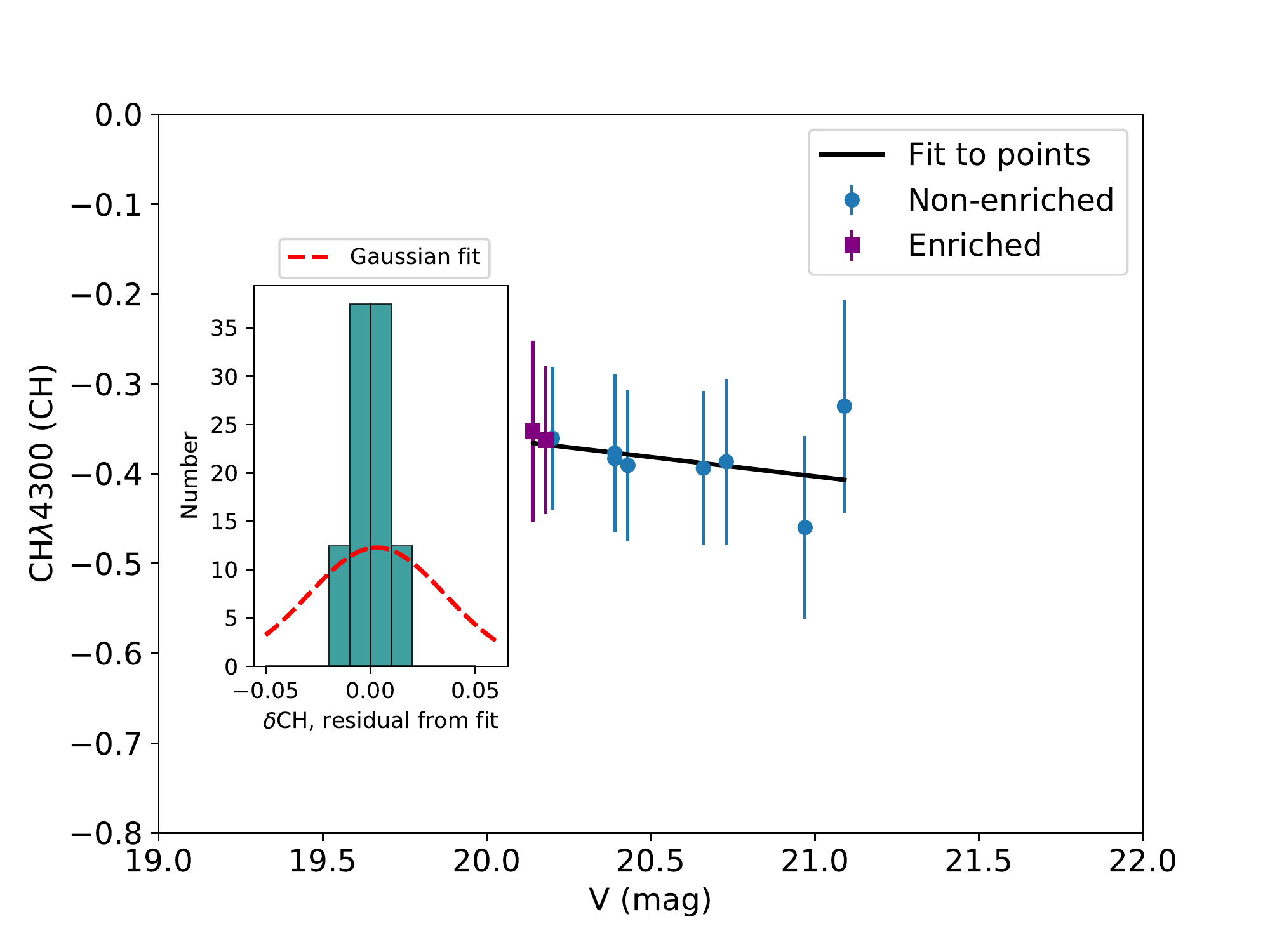}
	\includegraphics[width=8.5cm]{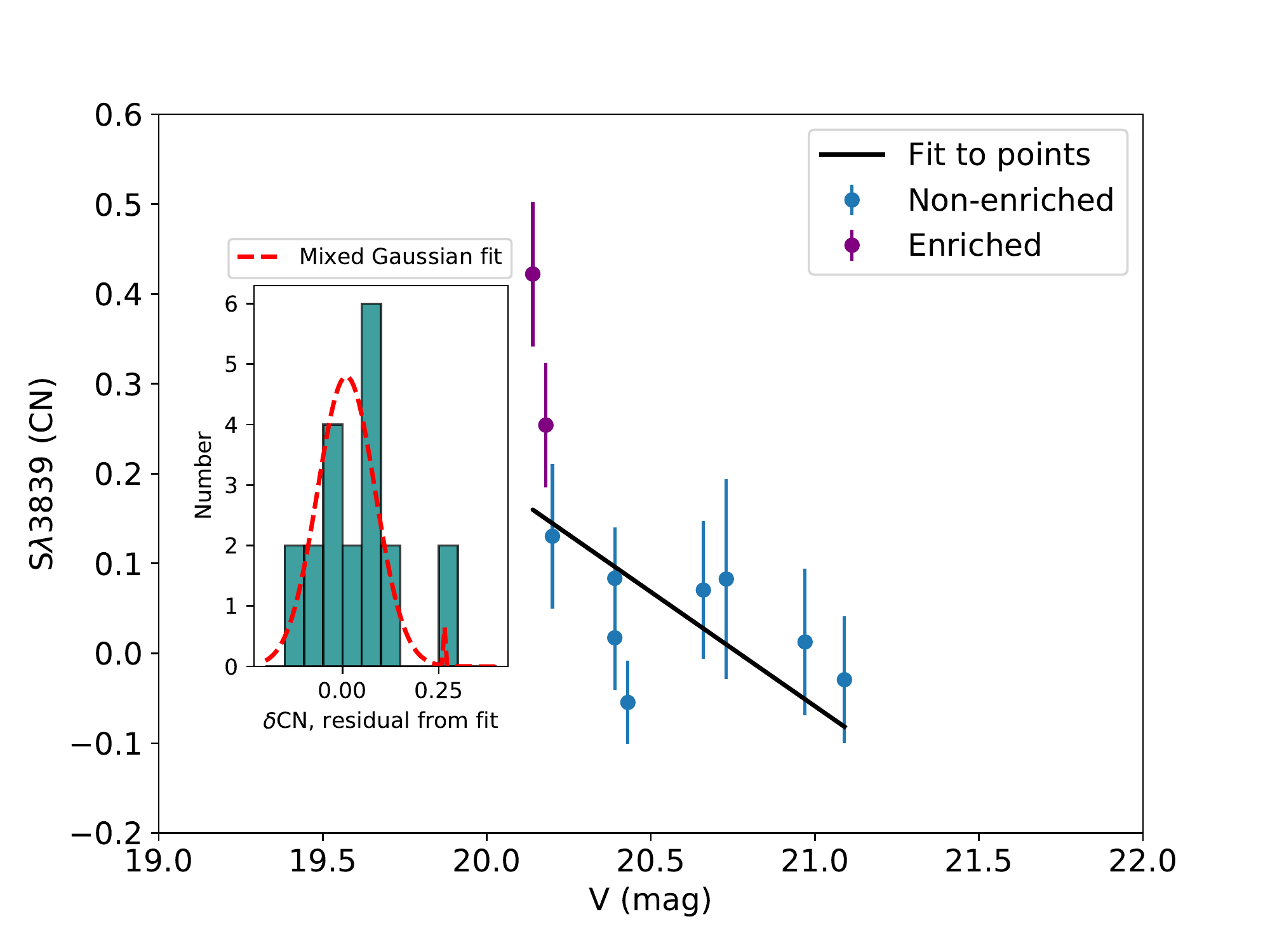}
	\caption{CH and CN plotted against V band magnitude for all member stars. N-enriched stars identified from the CN/CH plot are shown in purple. The data is fit in each case with a straight line (black line) and the inner plots show the residuals of each point from the fit. }
	\label{fig:gaussian}
\end{figure*}

This difference between CH and CN is also shown in Fig.~\ref{fig:gaussian}. Here we show CH and CN, respectively, against V band magnitude. Again, the axes cover the same range in order to illustrate the difference. We fit the points with a straight line (shown in black) and show the distributions of the residuals of the fit in the inner plots. The histograms are fitted with the {\sc mixed gaussian} routine in the {\sc sklearn} package in {\sc python}.

\begin{figure}
	\includegraphics[width=8.5cm]{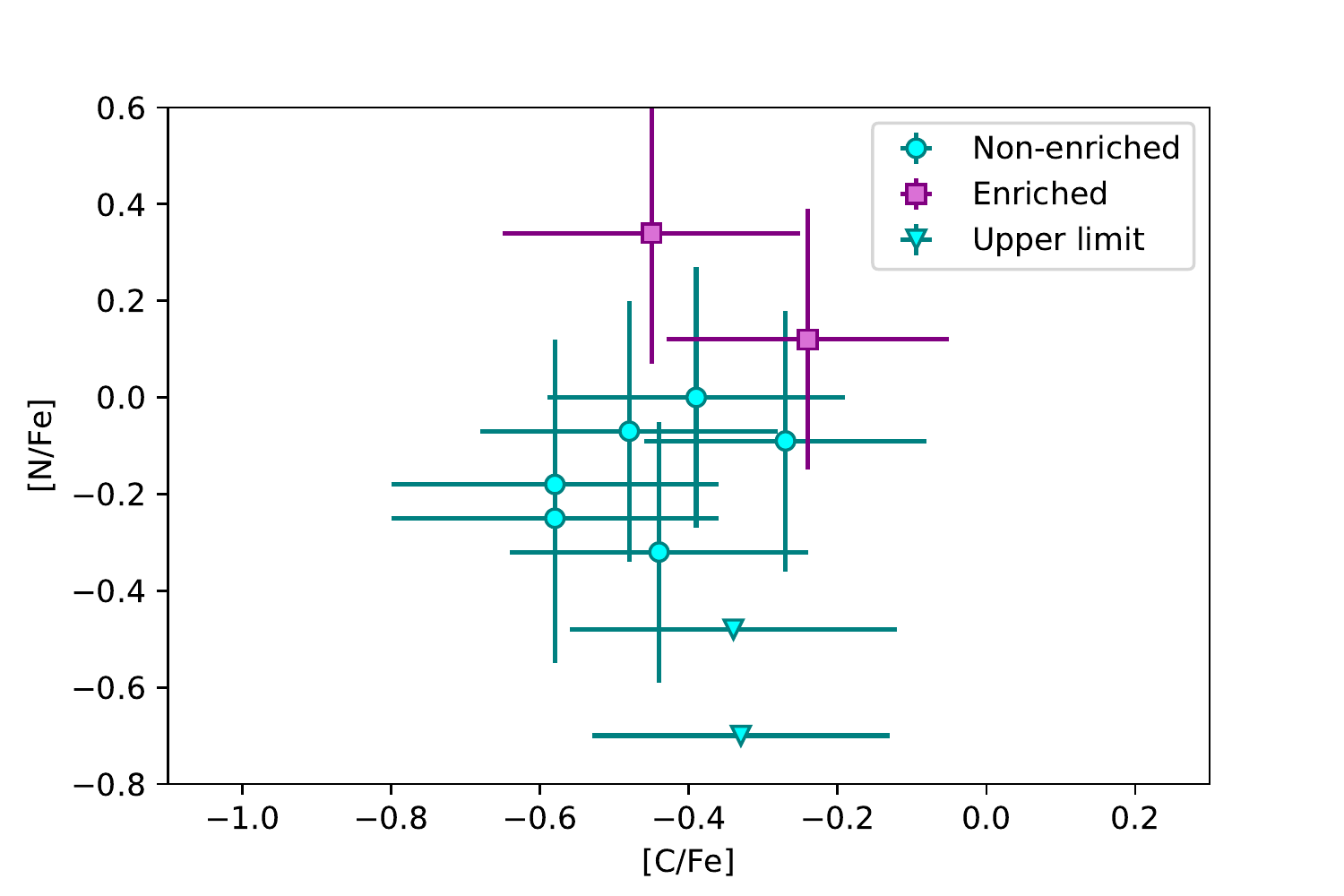}
	\caption{[C/Fe] and [N/Fe] for all member non-contaminated stars. As with CN and CH we see a spread in [N/Fe] without a spread in [C/Fe]. This shows that our stars enriched in CN are also enriched in [N/Fe] and therefore CN is representative of nitrogen abundance. The errors are very large on these estimations due to low resolution spectra and high amounts of noise. Purple squares are enriched stars (selected from CN/CH), blue circles are non-enriched stars and blue triangles are non-enriched stars but with only upper limits on the measurements of [N/Fe]. }
	\label{fig:cn}
\end{figure}

Fig.~\ref{fig:cn} shows our estimates of the [C/Fe] and [N/Fe] abundances for all non-contaminated member stars. Though the errors are very large and make it difficult to interpret any definite results, we can still roughly see a larger spread in [N/Fe] than [C/Fe], as mirrored in the CN/CH distribution. Additionally, the two purple points indicate the CN enriched stars selected from the CN/CH plot, which also have higher [N/Fe] than the other points. Therefore this indicates that CN does reliably trace nitrogen abundance.

The method using CN and CH band strengths to investigate MPs is reliable, as it shows the expected results for Milky Way GCs \citep{kayser08,pancino10} and our previous result with Lindsay~1 \citep{hollyhead17} using the same process was also confirmed with HST photometry \citep{niederhofer17}. The technique has an advantage over high resolution spectroscopy, which would be required for accurate [C/Fe] and [N/Fe] (though they can be estimated), in allowing for a larger sample of stars to be observed in less time. Looking for MPs requires a fairly large number of stars in order to ensure any enriched stars are not missed, as in some clusters the enriched population appears to be more centrally concentrated \citep{lardo11,simioni16,dalessandro16}. Unlike photometry, however, this method is limited by the central regions of the cluster that are too crowded to sample, meaning the ratio of enriched to non-enriched stars cannot reliably be estimated from spectroscopy. 

\begin{figure}
	\includegraphics[width=8.5cm]{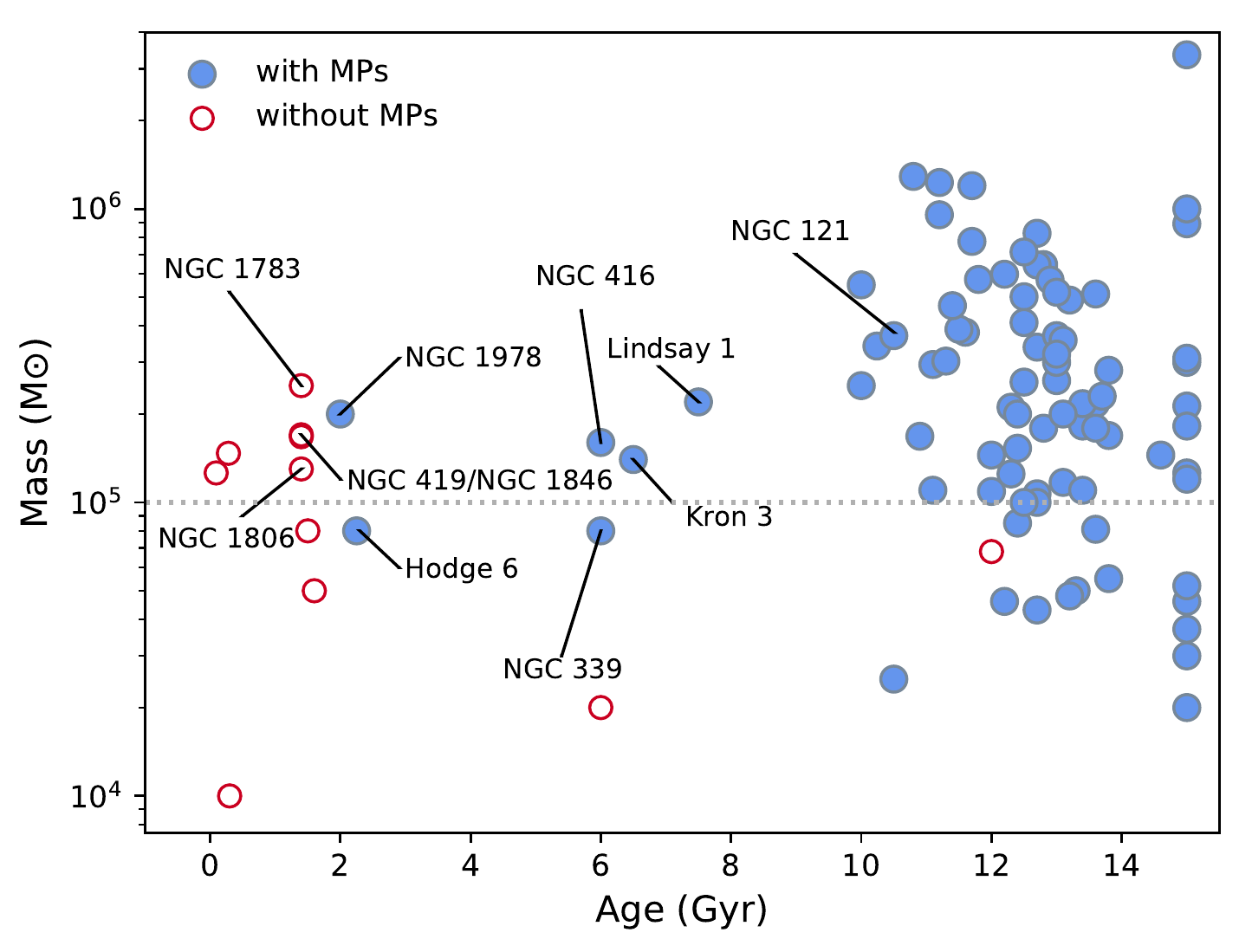}
	\caption{Age vs mass for all clusters studied so far looking for the presence of multiple populations. Filled circles indicate clusters where MPs are found and empty circles are those without MPs. The plot indicates that age plays a clear role in whether a cluster has MPs or not, with the transition from MPs to none occurring at the age at which the eMSTO phenomenon is no longer observed.}
	\label{fig:agemass}
\end{figure}

Fig.~\ref{fig:agemass} shows the relationship between cluster age, mass and the presence of MPs. This plot illustrates that MPs are present in all clusters of sufficient mass down to the age of $\sim$2~Gyr. Below this age, clusters show no spectroscopic or photometric evidence for MPs. The confirmation of the presence of MPs in clusters at $\approx$2~Gyr old such as Hodge~6 or NGC~1978 \citep{martocchia18b,martocchia18a} has significant consequences for GC formation theories, as it means the mechanism must still be operating in the present day.

This apparent age limit for MPs is also interesting due to its coincidence with the disappearance of eMSTOs, which are observed in clusters younger than $\sim$2~Gyr, but none older. Hodge~6 does not show a prominent eMSTO, as the width of the turn-off has been estimated as $\sigma$ < 100 Myr \citep{goudfrooij14}.

The apparent coincidence of the lack of eMSTOs and the beginning of MPs could well be just that - a coincidence. Relatively very few numbers of clusters have been studied at this exact age limit and so there is not the statistical consensus to indicate that these two phenomena are related, though studies so far do point to an age dependence on the onset of MPs.

Further study of clusters at this age limit is needed to determine whether these two phenomena are related. If eMSTOs can predict which young clusters will develop MPs, these objects can be used to determine the mechanism for the onset of MPs. Our result also suggests that YMCs can be considered analogues to GCs and used to determine their formation.
	
This coincidence could potentially be related to the mass of the stars on the RGB when abundances of these stars are measured. The $\approx$2~Gyr age limit also corresponds to the mass of RGB stars changing to lower mass stars of $\approx$1.5\msun. Therefore, the observation of MPs in clusters is related to the mass of the stars that are observed and less evolved stars would need to be studied in clusters younger than $\approx$2~Gyr. There is evidence, however, that there is an increase in the observability of MPs with age \citep{martocchia18a}, and $\approx$2~Gyr may be the youngest age where they are observable.

Finally, as said previously, the sample of clusters that are studied for this purpose is fairly small and constrained to the local universe. Improving our ability to study clusters to greater distances and therefore sampling a wider range of environments would be highly beneficial to interpreting these results and discovering the mechanism for the onset of MPs.


\section*{Acknowledgements}
Based on observations made with ESO telescopes at the La Silla Paranal Observatory under Programme ID 099.D-0762(B), P.I. K Hollyhead.

We thank P. Goudfrooij for making the HST photometry of Hodge 6 available to us.

C.~Lardo thanks the Swiss National Science Foundation for supporting this research through the Ambizione grant number PZ00P2 168065.

N.~Bastian gratefully acknowledges financial support from the Royal Society (University Research Fellowship) and the European Research Council (ERC-CoG-646928-Multi-Pop). 

F.~Niederhofer acknowledges support from the European Research Council (ERC) under European Union's Horizon 2020 research and innovation programme (grant agreement No 682115).

Support for this work was provided by NASA through Hubble Fellowship grant HST-HF2-51387.001-A awarded by the Space Telescope Science Institute, which is operated by the Association of Universities for Research in Astronomy, Inc., for NASA, under contract NAS5-26555. 

C.~Usher gratefully acknowledges financial support from European Research Council (ERC-CoG-646928-Multi-Pop).


\bibliographystyle{mnras}
\bibliography{biblio}

\bsp	
\label{lastpage}


\end{document}